\documentclass[journal]{IEEEtran}
\usepackage{color,array,amsthm}
\usepackage{cite}
\usepackage{amsmath,amssymb,amsfonts,graphicx}
\usepackage{algorithmic}
\usepackage{svg}
\usepackage{graphicx}
\usepackage{mathtools}
\usepackage{textcomp}
\usepackage{xcolor}
\usepackage{textcomp}
\graphicspath{ {./} }

\begin{document}
\title{Changepoint Detection for Real-Time Spectrum Sharing Radar}
\author{Samuel~Haug, Austin~Egbert, Robert~J.~Marks~II, Charles~Baylis, Anthony~Martone}





\maketitle

\begin{abstract}
Radar must adapt to changing environments, and we propose changepoint detection as a method to do so. In the world of increasingly congested radio frequencies, radars must adapt to avoid interference. Many radar systems employ the \textit{prediction action cycle} to proactively determine transmission mode while spectrum sharing. This method constructs and implements a model of the environment to predict unused frequencies, and then transmits in this predicted availability. For these selection strategies, performance is directly reliant on the quality of the underlying environmental models. In order to keep up with a changing environment, these models can employ \textit{changepoint detection}. Changepoint detection is the identification of sudden changes, or changepoints, in the distribution from which data is drawn. This information allows the models to discard ``garbage" data from a previous distribution, which has no relation to the current state of the environment. In this work, \textit{bayesian online changepoint detection} (BOCD) is applied to the sense and predict algorithm to increase the accuracy of its models and improve its performance. In the context of spectrum sharing, these changepoints represent interferers leaving and entering the spectral environment. The addition of changepoint detection allows for dynamic and robust spectrum sharing even as interference patterns change dramatically. BOCD is especially advantageous because it enables online changepoint detection, allowing models to be updated continuously as data are collected. This strategy can also be applied to many other predictive algorithms that create models in a changing environment.
\end{abstract}


\section{Introduction}
\IEEEPARstart{A}{s} more and more spectral users enter the environment, frequency bands exhibit more limited availability. Radars that operate efficiently in such congested environments must be able to overcome this sometimes considerable obstacle.

One technique to allow radars to operate normally in congested spectral environments is to create and maintain a model of the spectral environment over time so that the radar can predict available frequency bands and transmit only in the predicted availability.

As potential interferers enter and leave the spectral environment, spectrum sharing radars must adapt to these changes for optimal performance. For spectrum evaluation and prediction algorithms that create models of the environment, such models can be greatly improved with the implementation of changepoint detection. Changepoint detection is the identification of sudden changes, or changepoints, in the distribution from which data is drawn. With this knowledge of changepoint locations, models can discard data from an unrelated distribution and increase their prediction accuracy.

Spectral evaluation and prediction algorithms that employ this changepoint detection can exhibit additional robustness to sudden and drastic changes in spectral environment behavior. This can allow the radar user to perform more successful observations with their radar and to be more confident that that information is accurate and not distorted by interference.

Though this work focuses specifically on applying Bayesian online changepoint detection to the sense and predict system, changepoint detection can likewise be applied to other systems that maintain a model of the environment \cite{SpectrumSharing} or employ machine learning on observed environmental behavior \cite{SpectrumSharing1}. Wang et al. \cite{survey} give a survey of many spectrum sharing techniques and algorithms. Our work is unique because it employs changepoint detection for increased robustness to sudden and drastic changes in environmental behavior.

As an example, an alternative changepoint detection method has been applied to the detection of interference in cognitive radio \cite{othercp1}. Our work is distinctive from that application because we consider large scale changes in interference patterns for the purpose of live prediction while the other mentioned work considers small scale interference changes for the purpose of detection.

\section{Background}
\subsection{Sense and Predict}
\begin{figure}[htbp]
     \centering
     \includegraphics[width=.8\textwidth,angle=270]{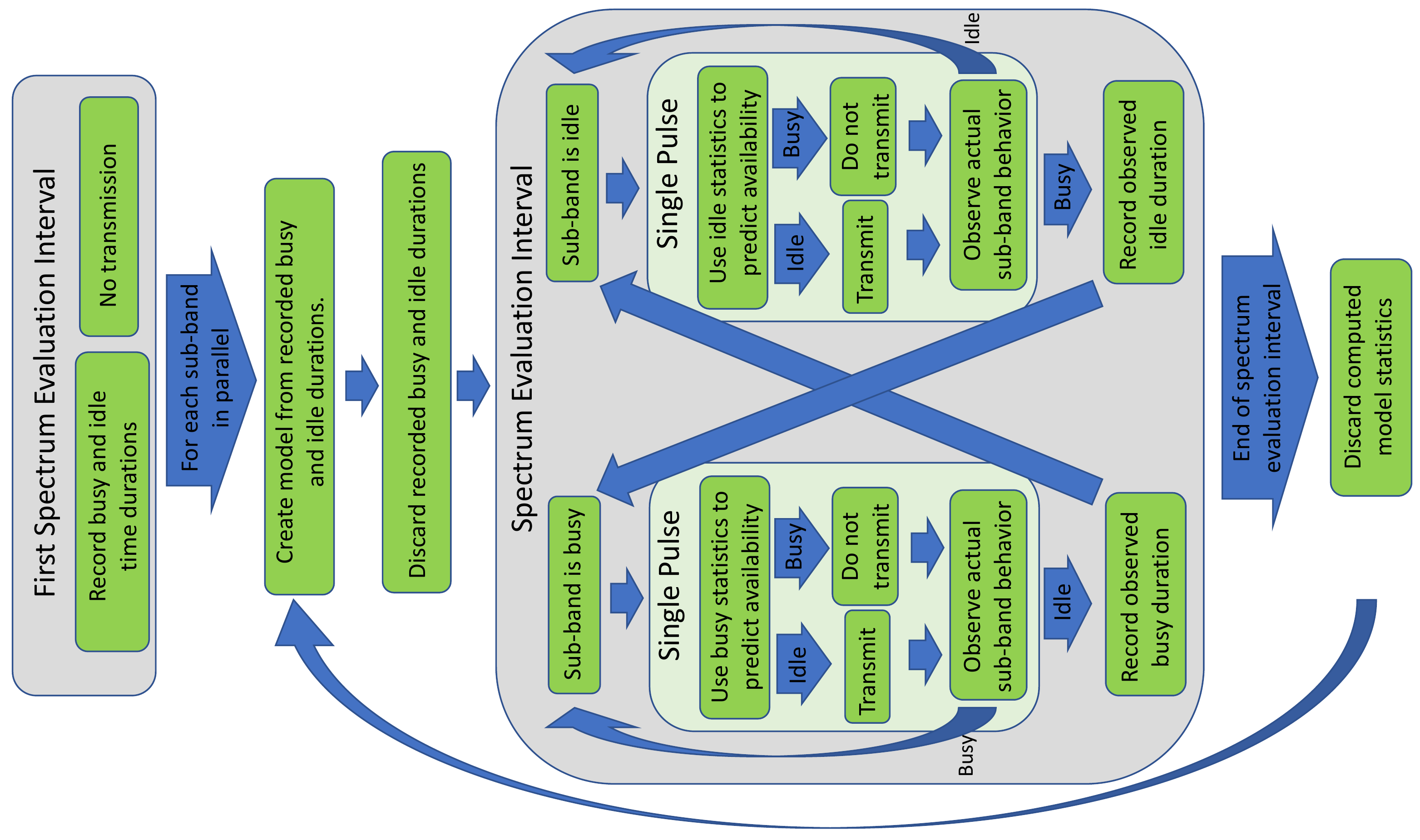}
     \caption{The sense and predict system. The length of the spectrum evaluation interval is user-defined and is a length of time, not a number of state transitions; that is, the quantity of individual busy and idle durations may vary during each evaluation interval.
     The transmission indicated is sub-band specific; lack of transmission in one sub-band does not mean that the radar as a whole ceases to function, but rather that the specific sub-band is avoided.}
     \label{fig:flow}
 \end{figure}

The sense and predict algorithm \cite{Kovarskiy} employs the prediction action cycle to inform intelligent radar frequency selection. In this model, the spectrum is broken up by frequency into sub-bands, with each sub-band considered as an independent \textit{alternating renewal process}. Busy and idle intervals are modeled using a log-normal distribution \cite{thes58, thes59}. In the initialization of the system, there is a period of passive spectral evaluation (training) before predictions and transmissions are made. When training is completed, model statistics are generated and used to inform predictions, and transmission begins. The time duration allocated for this training phase is denoted as a \textit{spectrum evaluation interval} (SEI). The data gathered in this SEI are used to create the model for the next SEI, and training data gathered in this second SEI is used to create the model for the successive SEI. In this way, the system uses training data from a given SEI to create the working model for the next SEI. The length of this SEI is chosen by the user. A block diagram of the sense and predict system is shown in Figure \ref{fig:flow}.

\begin{figure}[htbp]
    \centering
    \includegraphics[width=.5\textwidth]{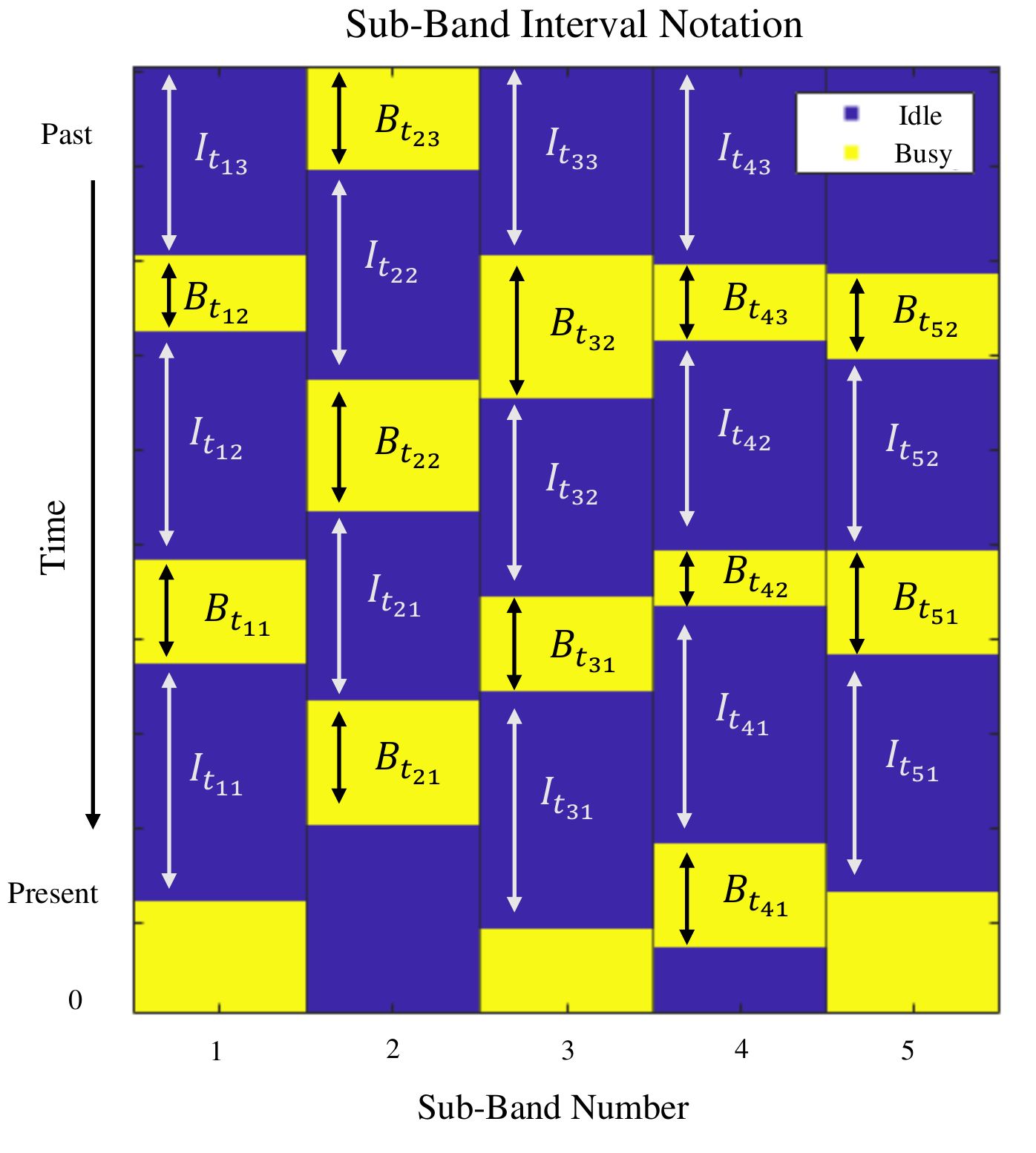}
    \caption{An illustration of sub-band behavior and model terminology. The present state of the environment is shown at time $=0$. The units of time are omitted because they vary widely by application. The length of time that a sub-band remains busy is referred to as a busy time duration or busy interval. The length of time that a sub-band remains idle is referred to as an idle time duration or idle interval. Note that busy and idle intervals that intersect with the present (at the bottom of the illustration) are unlabeled, because the system has not observed the end of those intervals and does not know how long each sub-band will remain in its current state. }
    \label{fig:ARP}
\end{figure}

The system uses a log-normal model to characterize the observed busy $B$ and idle $I$ time durations. The timeline is therefore divided up at each point as either being busy or idle. The length of time that a sub-band spends being both busy and idle is recorded, and two distributions are developed for both busy and idle time durations. An illustration of sub-band behavior and notation is shown in Figure \ref{fig:ARP}.
The system's model of the environment therefore includes two probability distributions for each sub-band. It then uses each sub-band's two parametric probability distributions to predict its future behavior. To determine distribution parameters, a sample mean estimate $\mu_B$ and variance estimate $\sigma^2_B$ are first computed for busy times in all $M$ sub-bands

$$\mu_{B_i}=\dfrac{1}{n_i}\sum^{n_i}_{j=1}B_{t_{ij}}$$
$$
\sigma^2_{B_i}=\dfrac{1}{n_i} (B_{t_{ij}}-\mu_{B_i})^2.
$$
The index of each frequency sub-band is denoted by $i\in\{ 1, 2, 3,...,M\}$. The number of observed busy intervals in the $i$-th band is $n_i$. The number of observed idle intervals is also $n_i$. The duration of the $j$-th busy interval in channel $i$ is $B_{t_{ij}}$. The statistics of the time process are assumed to be log-normal with a mean of $\hat{\mu}_{B_i}$ and a standard deviation of $\hat{\sigma}_{B_i}$. From this assumption the computed sample statistics are log-normal parameters. The logarithm of a log-normal distribution is a Gaussian distribution with mean $\mu$ and variance $\sigma^2$. For the $i$-th sub-band busy time durations the corresponding parameters are $\mu_{B_i}$ and $\sigma^2_{B_i}$. The log-normal mean and standard deviation can be computed as a function of the Gaussian mean and standard deviation. Specifically,
\begin{equation}\label{log1}
\hat{\mu}_{{B_i}}=\ln\left(\dfrac{\mu_{B_i}^2}{\sigma_{B_i}^2+\mu_{B_i}^2}\right)
\end{equation}
\begin{equation}\label{log2}
 \hat{\sigma}_{{B_i}}=\sqrt{\ln\left(\dfrac{\sigma_{B_i}^2}{\mu_{B_i}^2}\right)+1}.
 \end{equation}
The idle mean and variance are calculated in the same way except $B$ subscripts are replaced by $I$ subscripts.

The system uses these calculated parameters to compute availability probabilities for each sub-band at every time-step in the next spectrum evaluation interval. These availability probabilities are calculated by referencing the distribution's \textit{cumulative distribution function} (CDF). The CDF for a lognormal distribution is
\begin{equation}
\label{logcdf}
\textrm{cdf}_L(\hat{\mu},\hat{\sigma},t)=\dfrac{1}{\hat{\sigma}\sqrt{2\pi}}\int_0^t \dfrac{1}{\tau}\exp\left(-\dfrac{(\ln{\tau}-\hat{\mu})^2}{2\hat{\sigma}^2}\right) d\tau.
\end{equation}
This CDF is used to determine the likelihood of a state change in a busy interval
\begin{equation} \label{p1}
p_{B_i}(t_{B_i}+\Delta t)=\textrm{cdf}_L(\hat{\mu}_{B_i},\hat{\sigma}_{B_i},t_{B_i}+\Delta t) 
\end{equation}
\begin{equation} \label{p2}
 p_{I_i}(t_{I_i}+\Delta t)=1-\textrm{cdf}_L(\hat{\mu}_{I_i},\hat{\sigma}_{I_i},t_{I_i}+\Delta t)
 \end{equation}
where $p_{I_i}(t+\Delta t)$ and $p_{B_i}(t+\Delta t)$ are the probabilities that the sub-band will be available after $\Delta t$ time-steps if the sub-band has already been idle or busy, respectively, for $t$ time-steps. When required, this equation allows for a system latency of multiple time-steps between observation and transmission. If there is no system latency, $\Delta t$ is set to a single pulse duration.

At each time-step after availability probabilities are calculated for each sub-band, a threshold $\theta$ is applied to predict the set of available sub-bands $\textbf{A}=\{A_1,A_2,A_3,...,A_M\}$. Because busy and idle intervals are modeled and evaluated separately, separate thresholds are used to evaluate availability probabilities of busy and idle sub-bands 
\begin{equation}\label{threshold}
A_i=\begin{dcases}
p_{B_i}\geq\theta_B, & S_i=1\\
p_{I_i}\geq\theta_I, & S_i=0.
\end{dcases}
\end{equation}
where $\theta_B$ represents the probability threshold for busy statistics, $\theta_I$ represents the probability threshold for idle statistics
$\theta_I,\theta_B \in [0,1]
$, and $S_i$ represents the current status of the $i$-th sub-band, with a one representing busy and a zero representing idle.

If the availability probability is above the threshold, the system will predict that sub-band to be available and attempt to transmit in that sub-band. If the availability probability is below the threshold, the system will predict it to be unavailable and will not attempt to transmit in that sub-band. These thresholds control a direct trade off between false alarms and missed detections with regard to predicted state change. A false alarm in this context signifies that the system believes a sub-band to be available when it is not, which results in a collision with interference. Missed detections in this context signify that the system predicted a sub-band to be unavailable when it is available, which results in a missed opportunity for transmission. Higher values for both $\theta_B$ and $\theta_I$ mean that the system requires more evidence to believe that a sub-band will be available, and will therefore cause more missed opportunities and fewer collisions. Conversely, lower values will have the opposite effect, causing more collisions and fewer missed opportunities.

Availability probability thresholds are optimized when sub-band models are created by using a grid search to exhaustively test 100 threshold values for both $\theta_I$ and $\theta_B$ between 0.05 and 0.95. The error rates of each threshold combination are then compared and the threshold combination that yields the least total weighted error $\rho$ is chosen and implemented. This total weighted error is calculated by determining the rate of collisions $C$ and missed opportunities $D$, and then implementing a convex weighting factor $\alpha $ between these two error metrics
\begin{equation}\label{rho}
\rho=\alpha C +(1-\alpha )D 
\end{equation}
where $0\leq \alpha \leq 1$.

\subsection{Bayesian Online Changepoint Detection}
\label{00}
\begin{figure}[htbp]
    \centering
    \includegraphics[width=.5\textwidth]{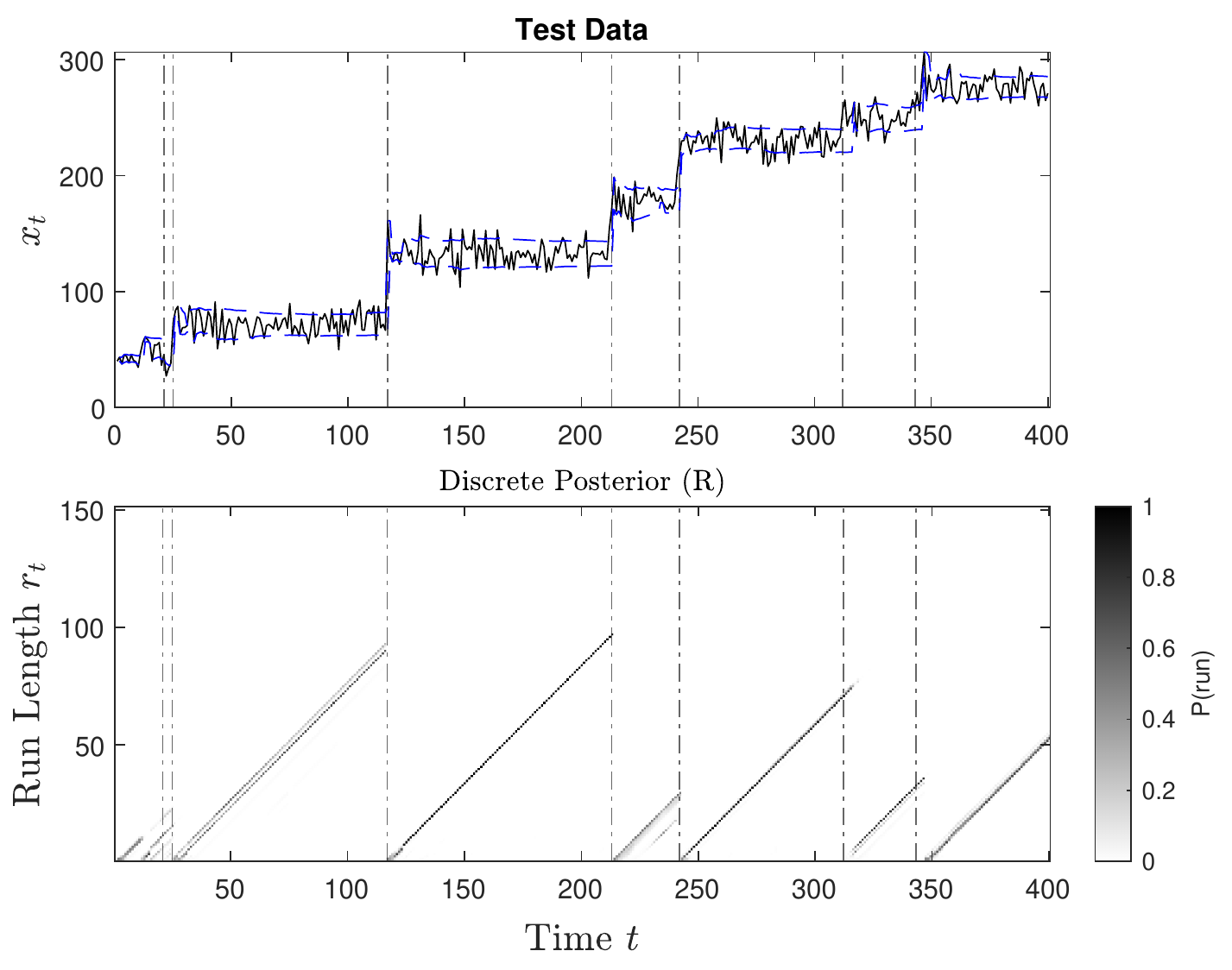}
    \caption{Discrete posterior generated by BOCD on arbitrary test data, with ground truth changepoints represented by vertical lines. Note that the algorithm correctly identifies changepoints by dropping the run length $r_t$ to zero at each vertical line.
    }
    \label{fig:my_label}
\end{figure}

While many changepoint detection algorithms take place offline after all data has been collected \cite{offline1, offline2, offline3, offline4}, Bayesian online changepoint detection (BOCD) can be implemented online \cite{BOCD}. BOCD employs causal predictive filtering to divide a data set into discrete partitions. All data in a partition are treated as i.i.d. A datum in the present partition at time $t$ is denoted by $x_t^{(r)}$. The variable $r_t$ represents the time since the last changepoint. Large values of $r_t$ indicate that the current partition is long and a changepoint has not occurred recently. As data is introduced to the system, all possible run lengths are considered, and each of these possible run lengths is assigned a probability $P(r_t|x_{1:t})$. Upon initialization in a new environment the first datum must be the first in its run length, and so $P(r_1=0|x_1)=1.$ As time increases there are two options: either the run length increases by one (no changepoint) or the run length remains at zero (changepoint). The probability that the run length will increase is called the \textit{growth probability}, and the probability that the run length will drop to zero is called the \textit{changepoint probability}. As more data are collected, probabilities for every combination of changepoints are calculated, creating a stochastic process called the \textit{discrete posterior}. The discrete posterior models discrete time $t$ in one dimension and the \textit{probability mass function} (PMF) for $r_t$ in the other dimension. Online evaluation is performed recursively
$$P(r_t,x_{1:t})=\sum_{r_{t-1}} P(r_t|r_{t-1})P(x_t|r_{t-1},x_t^{(r)})P(r_{t-1},x_{1:t-1}) $$
where 
$$P(r_t|r_{t-1}) =\begin{dcases}
H(r_{t-1}+1), & r_t=0 \\
1-H(r_{t-1}+1), & r_t=r_{t-1}+1 \\
0, & \textrm{otherwise} \\
\end{dcases}$$
and $H(t)$ is the \textit{hazard function}. In the case of a memoryless hazard function such as an exponential decay, $H(t)$ is a constant $h$. This recursive operation does not produce a normalized PMF, and so the resulting mass function must be normalized to have a total value of 1. $ P(x_t|r_{t-1},x_t^{(r)})$ represents the probability of generating the newest datum using a model of the environment for run length $r_t$, represented below by $\pi_t^{(r)}.$ An example of the discrete posterior for a time series in the $x$ domain is shown in Figure \ref{fig:my_label}.

BOCD uses a Bayesian prior to help inform its model $\pi_t^{(r)}$, which requires \textit{a priori} knowledge of the upcoming data series, including probability distribution hyperparameters. BOCD originally required an estimate of the changepoint prior, which is the changepoint frequency $h$ of future data.

Later versions of this technique allow for online estimation of this changepoint prior with unknown changepoint locations, which reduces the amount of \textit{a priori} information that is required \cite{hazard}. This system uses a joint probability mass function of $r_t$ and $a_t$, where $a_t$ is the number of changepoints that have occurred since $t=0$. In this joint distribution a similar method is used to determine changepoint and growth probabilities, though any instances of hypothesized changepoints not only reset $r_t$ to zero, but $a_t$ also increases by one. A time-step with no changepoint increases $r_t$ by one and $a_t$ remains the same. The joint probability mass function is calculated recursively as
        \begin{align}
                 P(r_t,a_t|x_{1:t}) &= \sum_{r_{t-1}}\sum_{a_{t-1}}P(r_t,a_t|r_{t-1},a_{t-1}) \nonumber  \\
                    & \times P(x_t|x_{t-1}^{(r_{t-1})})P(r_{t-1},a_{t-1}|x_{1:t-1})\notag
        \end{align}
where 
$$ P(r_t,a_t|r_{t-1},a_{t-1})$$ $$=
\begin{dcases}
\dfrac{b_{t-1}+1}{a_{t-1}+b_{t-1}+2}=1-\tilde{h}_t,  & r_t=r_{t-1}+1 \ \& \
a_t=a_{t-1} \\ 
\dfrac{a_{t-1}+1}{a_{t-1}+b_{t-1}+2}=\tilde{h}_t, &  r_t=0 \ \& \ a_t=a_{t-1}+1 \\
0, & \textrm{otherwise}
\end{dcases}$$
and $b_t$ is a count of how many nonchangepoints have occurred since time $t=0$, given by $b_t=t-a_t$. This joint probability mass function is used to estimate the changepoint prior.
\section{BOCD Alterations}
In order for the BOCD algorithm to be compatible with the sense and predict algorithm, several alterations are made. Firstly, the system will be required to analyze an environment for which it has no prior information. Secondly, the system will be required to perform adequately on large amounts of data.
\subsection{Removing the Need for \textit{A Priori} Information}
As mentioned in Section \ref{00}, to inform Bayesian priors, several pieces of information regarding future distributions are required before any data are collected. In an effort to reduce the risk of an incorrect prior in a completely unknown environment, these priors are not employed, and model parameters are drawn entirely from collected data. The underlying probabilistic model $\pi_t^{(r)}$ for a run length of $r$ is generically chosen to be a normal distribution with parameters $\mu^{(r)}$ and $\sigma^{(r)}$. Because $r$ is the run length, it is also the number of data in consideration.

The mean of a run length is determined by averaging the observed values in a given partition
$$\mu^{(r)}=\dfrac{1}{r}\sum_{i=0}^{r-1} x_i
$$
where $x_0$ is the most recent observed data point. Similarly, the variance of the data is calculated using the first and second moments of the data in the partition
$$(\sigma^{(r)})^{2}=\dfrac{1}{r}\sum_{i=0}^{r-1} x_i^2 -(\mu^{(r)})^2.$$
As hypothesized run lengths of a single observation do not yield a useful standard deviation, all run lengths are assumed to contain at least two data
$$P(r_t|x_{1:t}) $$ 
$$=\begin{dcases}
\sum_{r_{t-1}\neq 0} h(\pi_t^{(r)})P(r_{t-1}|x_{1:t-1}), &  r_t=0 \\
P(r_{t-1}|x_{1:t-1}), &  r_t=r_{t-1}+1=1\\
(1-h)\pi_t^{(r)}P(r_{t-1}|x_{1:t-1}), & r_t=r_{t-1} +1\neq 1 \\
0, & \textrm{otherwise.} 
\end{dcases}
$$
Thus, the standard deviation of a new partition is initially scaled to the first two data in that partition. As more data are collected, this hypothesis is either supported or rejected.

As a result of this lack of \textit{a priori} information, a tuning parameter $\gamma \in (0,\infty)$ has been added to control the sensitivity of the algorithm

$$P(r_t|x_{1:t}) $$ 
$$=\begin{dcases}
\sum_{r_{t-1}\neq 0} h(\pi_t^{(r)})P(r_{t-1}|x_{1:t-1}), &  r_t=0 \\
P(r_{t-1}|x_{1:t-1}), &  r_t=r_{t-1}+1=1\\
(1-h)\pi_t^{(r)}P(r_{t-1}|x_{1:t-1})\gamma, & r_t=r_{t-1} +1\neq 1 \\
0, & \textrm{otherwise.} 
\end{dcases}
$$
Higher values of $\gamma$ favor growth probabilities and will less often detect changepoints.

In addition to removing these priors, online estimation of the hazard rate is employed \cite{hazard}, as summarized in Section \ref{00}.

\subsection{Computational Cost Improvements}
The original algorithm's computational complexity increases linearly with time, which eventually becomes infeasible. To combat this run length probabilities in the tail of the distribution with probabilities less than a threshold $\theta_r$ are discarded \cite{BOCD}. While this does reduce the computational cost, the overall linear trajectory of the computational cost is unchanged. To limit the worst-case cost of the algorithm to a constant level, a run length maximum $L$ is imposed. Probability mass that would move above this limit in the discrete posterior is instead accumulated at $r_t=L$
$$P'(r_t=L|x_{1:t})=P(r_t=L|x_{1:t})+\sum_{k=L+1}^{\infty}P(r_t=k|x_{1:t})$$ 
$$P'(r_t=k|x_{1:t})=0\textrm{ } \forall \textrm{ } k \in [L+1, \infty].$$
This alteration to the algorithm does not affect its ability to detect changepoints, though it does effectively limit the history window of the algorithm to only consider a maximum of $L$ data. This data truncation does not significantly alter the accuracy of the system. For run length estimations less than $L$, estimations are not affected at all. For run length estimations greater than $L$, the data truncation should not have a significant effect on the quality of the models because all of those data are assumed to be i.i.d. The value of $L$ directly controls a trade-off between decreasing computational cost and increasing the accuracy of the changepoint detection.

\begin{figure}[htbp]
    \centering
    \includegraphics[width=.5\textwidth]{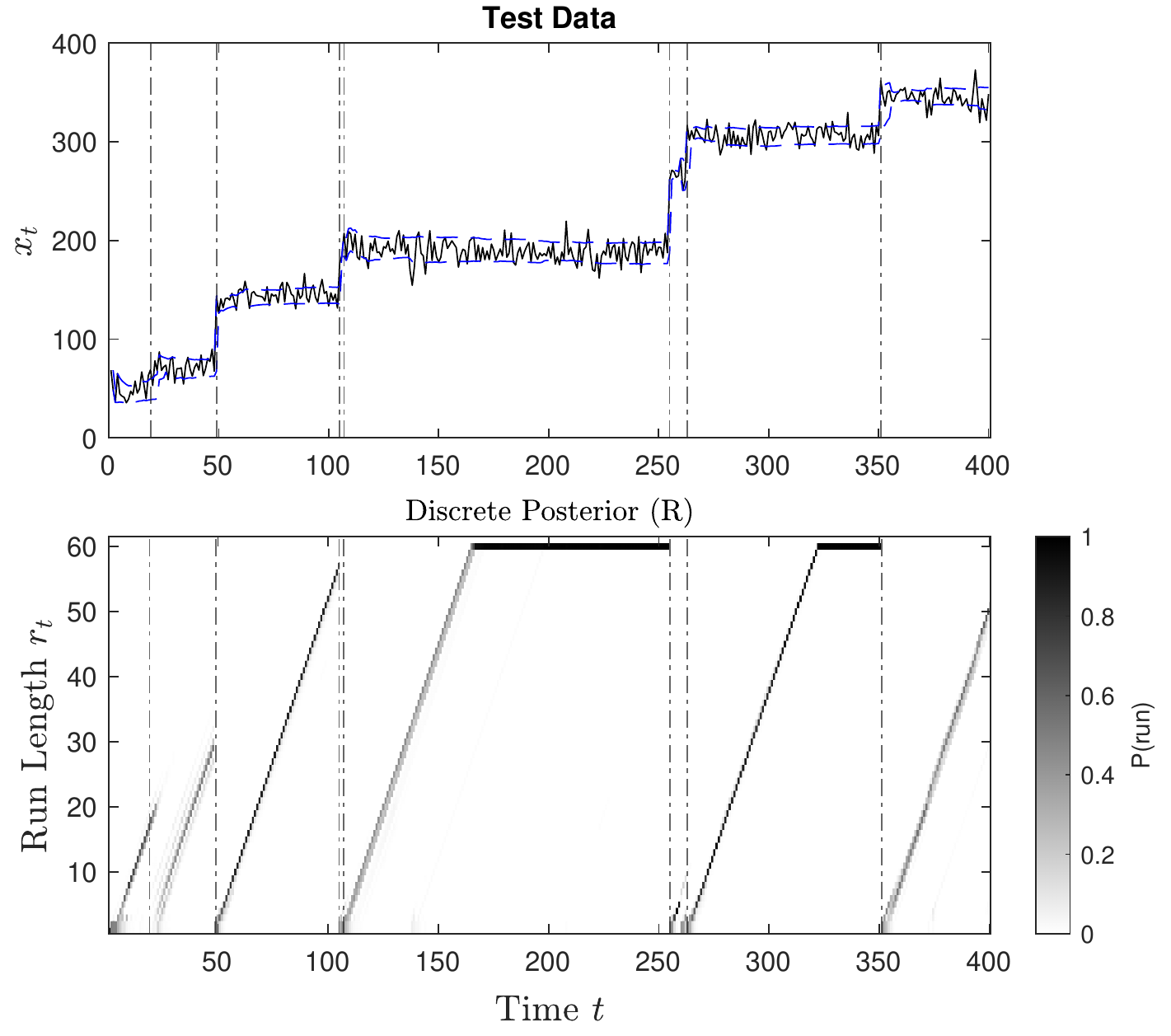}
    \caption{Discrete posterior with maximum run length $L=60$. Ground truth changepoints represented by vertical lines. Note the accumulation of probability mass at $L$ for large run lengths.}
    \label{fig:my_label2}
\end{figure}

An example of BOCD with an imposed run length maximum is shown in Figure \ref{fig:my_label2}.
In order to reduce the number of calculations required to maintain a joint probability mass function of both $r_t$ and $a_t$, probability masses in $a_t$ lower than a given threshold $\theta_a$ at either tail of the distribution are also discarded.
\section{Sense and Predict Alterations}
\subsection{Conditional Failure Probability}
Rather than using the value of a probability distribution's CDF to determine a sub-band's change of state probability, a conditional failure probability is employed. The conditional failure probability of a probability distribution is the probability that a failure will occur in an interval given no failure before that time. For any generic probability distribution $X$ this probability is given by
$$
P(a \leq X \leq b \text{ }| X \geq a)=\dfrac{\int_a^b f_X(\tau)d\tau}{\int_a^\infty f_X(\tau)d\tau}
$$
where $f_X(x)$ is the distribution's \textit{probability density function} (PDF). Using the relationship between a distribution's PDF and its CDF, this can be rewritten as
$$P(a \leq X \leq b \text{ }| X \geq a)=\dfrac{F_X(b)-F_X(a)}{1-F_X(a)} 
$$
where $F_X(x)$ is the distribution's CDF. This metric is generic and can be applied to any distribution.

This method is used to determine the availability probabilities for the $i$-th sub-band
\begin{equation}
\label{busy_availability_prob}
p_{B_i}(t_{B_i},\Delta t) =\dfrac{F_{B_i}(t_{B_i}+\Delta t)-F_{B_i}(t_{B_i})}{1-F_{B_i}(t_{B_i})}
\end{equation}
\begin{equation}
\label{idle_availability_prob}
p_{I_i}(t_{I_i},\Delta t) =1-\dfrac{F_{I_i}(t_{I_i}+ \Delta t) -F_{I_i}(t_{I_i})}{1-F_{I_i}(t_{I_i})}
\end{equation}
where $t_{B_i}$ and $t_{I_i}$ is the current length of the busy or idle interval respectively, $F_{B_i}$ and $F_{I_i}$ are the sub-band's computed lognormal busy and idle interval distribution CDF's respectively
$$
F_{B_i}(t)=\textrm{cdf}_L(\hat{\mu}_{B_i},\hat{\sigma}_{B_i},t)
$$
$$
F_{I_i}(t)=\textrm{cdf}_L(\hat{\mu}_{I_i},\hat{\sigma}_{I_i},t),
$$
 and $\Delta t$ is the duration of the action latency period, in units of radar pulse repetition intervals.

The set of available sub-bands \textbf{A} is then predicted by applying \eqref{threshold}.

\subsection{Nonparametric Model of Interval Data}
Though frequency band busy and idle interval periods are often modeled using a log-normal distribution, nonparametric modeling of these distributions has advantages.
In the nonparametric technique of modeling, a distribution's PMF is simply a histogram of observed data normalized to have total area of one. A cumulative sum of this PMF provides the distributions's CDF
$$
\textrm{cdf}_E(t)=\sum_{i=1}^t\textrm{pmf}_E(i)
$$
where $\textrm{cdf}_E(t)$ is the empirical CDF and $\textrm{pmf}_E(t)$ is the empirical PMF.
The conditional failure probabilities are then calculated with \eqref{busy_availability_prob} and \eqref{idle_availability_prob} using each sub-band's empirical CDF.

\begin{figure}[htbp]
    \centering
    \includegraphics[width=.5\textwidth]{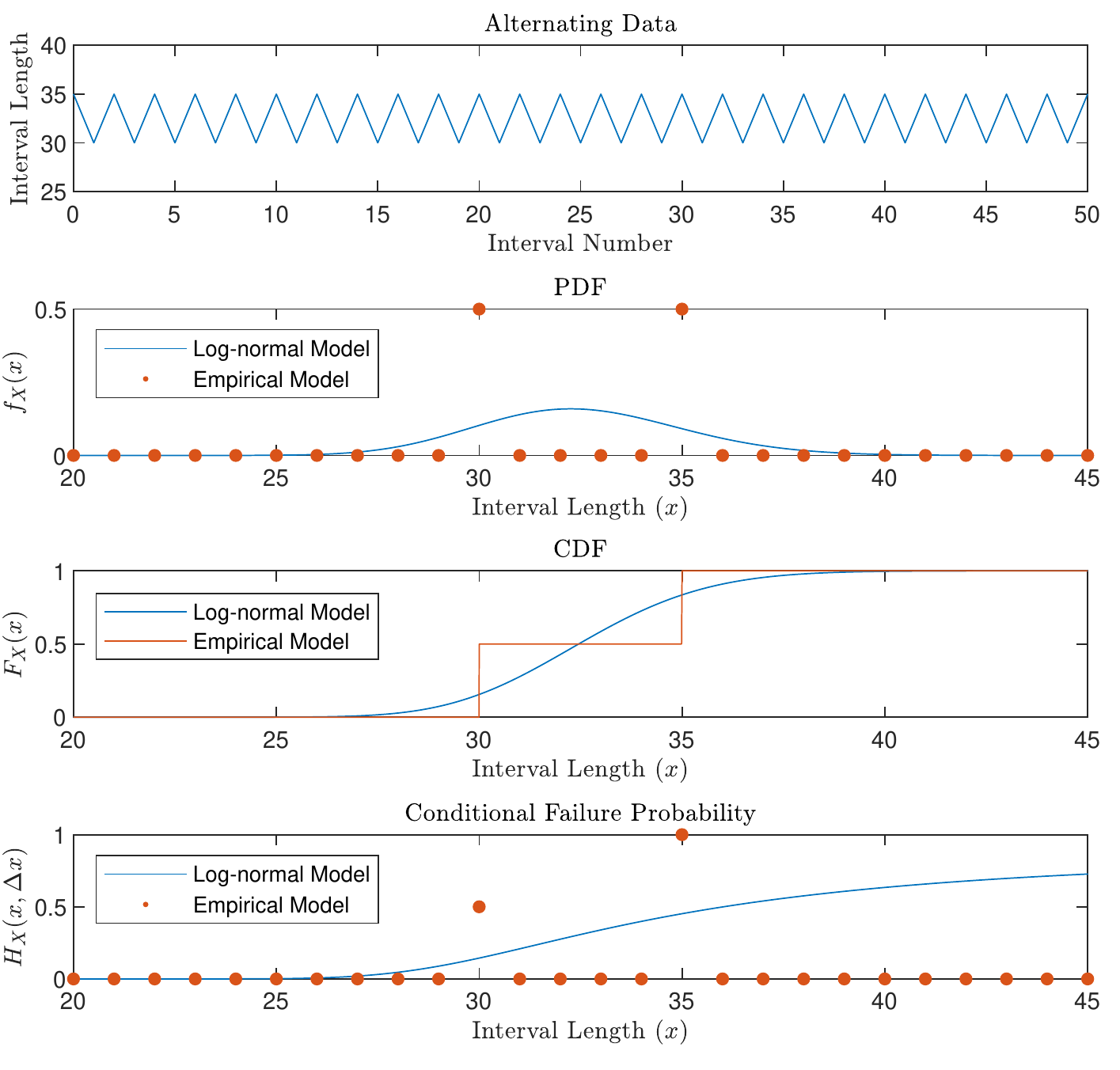}
    \caption{Pattern recognition with nonparametric model, $\Delta t = 1.$ To emphasize pattern recognition, very predictable alternating data is chosen. Note that the log-normal model is unable to reflect pattern recognition in conditional failure probability peaks at the appropriate interval lengths.}
    \label{fig:my_label3}
\end{figure}

One significant advantage of empirically modeling data is large-scale pattern recognition. A parametric model of data is unable to pick up on patterns such as repeated alternation between two values. The model simply scales a parametric curve as best it can to fit the data, but the underlying pattern is lost to the model. A nonparametric model however, can identify and adapt to such patterns. An example of this nonparametric pattern recognition is shown in Figure \ref{fig:my_label3}.

This pattern recognition can also lead to poor predictions in certain circumstances. During periods shortly after a changepoint has occurred in a data series, the model does not have many observed interval lengths. In this situation, the model will not predict any run lengths other than those that have already been observed. A parametric model could potentially yield better predictions in cases like these with little data and an environment that does conform well to a parametric model. In addition, a widely distributed uniform PMF produced by this nonparametric modeling may lead to relatively poor predictions when converted to conditional failure probabilities and thresholded.

Until both the lognormal and nonparametric methods of modeling are implemented and their relative performance is tested in a realistic environment, the model type employed by the system is a user defined parameter.
\section{Algorithm Integration}
In the overarching sense and predict system, the BOCD algorithm is employed to provide accurate and current estimates of busy and idle interval times $\mu_{B_i}$, $\sigma_{B_i}$, $\mu_{I_i}$, and $\sigma_{I_i}$ for each sub-band. Two BOCD models are allocated for each of the $M$ sub-bands, one to determine idle interval statistics, and one to determine busy interval statistics.  As the end of a sub-band's idle or busy interval is observed, the length of that interval is fed into the appropriate BOCD model as its $x_t$, and the joint probability mass function is updated. After this update, the run length $r_t$ with the highest probability is selected and the sample mean $\mu^{(r)}$ and variance $(\sigma^{(r)})^{2}$ of the corresponding run length is used as parameters $\mu_{I_i}$ and $\sigma^2_{I_i}$ for idle interval models and parameters $\mu_{B_i}$ and $\sigma^2_{B_i}$ for busy interval models.

If the system is employing the lognormal method of modeling, these statistics are then used to compute log-normal distribution parameters $\hat{\mu}_{B_i}$, $\hat{\sigma}_{B_i}$, $\hat{\mu}_{B_i}$, and $\hat{\sigma}_{B_i}$ using \eqref{log1} and \eqref{log2}. After this conversion, log-normal parameters are used to calculate availability probabilities for each sub-band using \eqref{busy_availability_prob} and \eqref{idle_availability_prob}. 

If the system is instead employing the nonparametric method of modeling, these statistics are used to generate both an empirical PDF and an empirical CDF. These are used to calculate the availability probabilities for each sub-band, also using \eqref{busy_availability_prob} and \eqref{idle_availability_prob} using the empirical CDF.

\begin{figure}[htbp]
    \centering
    \includegraphics[width=.8\textwidth,angle=270]{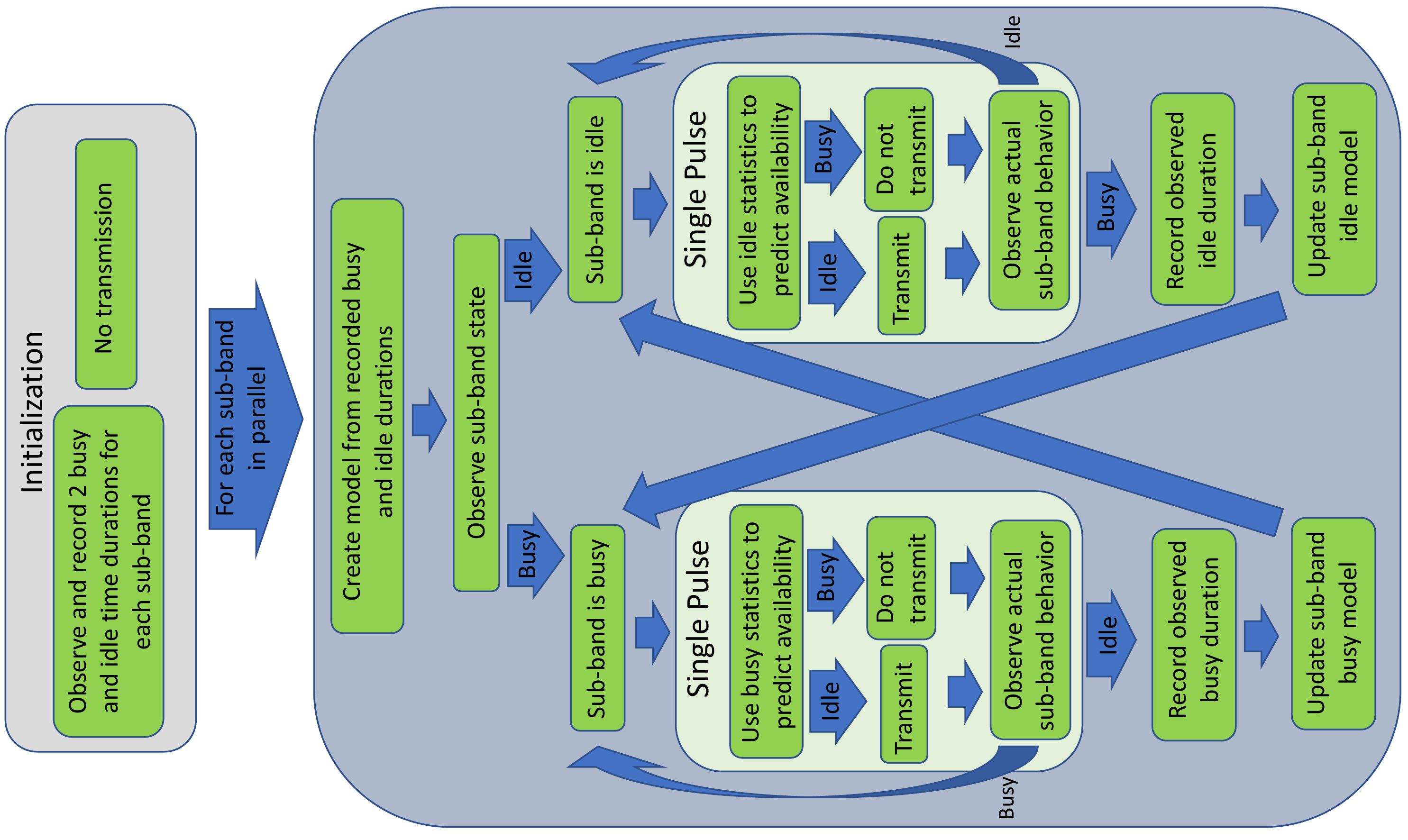}
    \caption{An updated system diagram for the sense and predict algorithm with changepoint detection. Note the absence of discrete spectrum evaluation intervals because learning is continuous.}
    \label{fig:flow_chart2}
\end{figure}

In this system an initial passive spectrum evaluation interval is not required. Interval distribution parameters are initially determined after two idle and busy intervals are observed for each sub-band. In similar fashion later models of the environment do not require a subsequent training interval because learning is continuous. Early models of the environment are based on very few data and may yield relatively poor predictions. As more intervals are observed, the system models the environment with as much data as is available. An updated system diagram is shown in Figure \ref{fig:flow_chart2}.
\subsection{Offline Threshold Determination}
Due to the frequent updates to all model parameters exhaustive, online threshold optimization with a grid-search is not feasible. As a result, threshold optimization are determined offline in advance. Optimal thresholds vary widely based on various environmental parameters, resulting in generic threshold selection
$$\theta_I=\theta_B=1 - \alpha $$
where $\alpha$ is the error weight factor used in \eqref{rho}. Higher collision weightings will yield optimal thresholds that require more evidence to predict a band to be available. Conversely lower collision weighting will have the inverse effect.

\section{Simulation Results}
\begin{figure}[htbp]
    \centering
    \includegraphics[width=.5\textwidth]{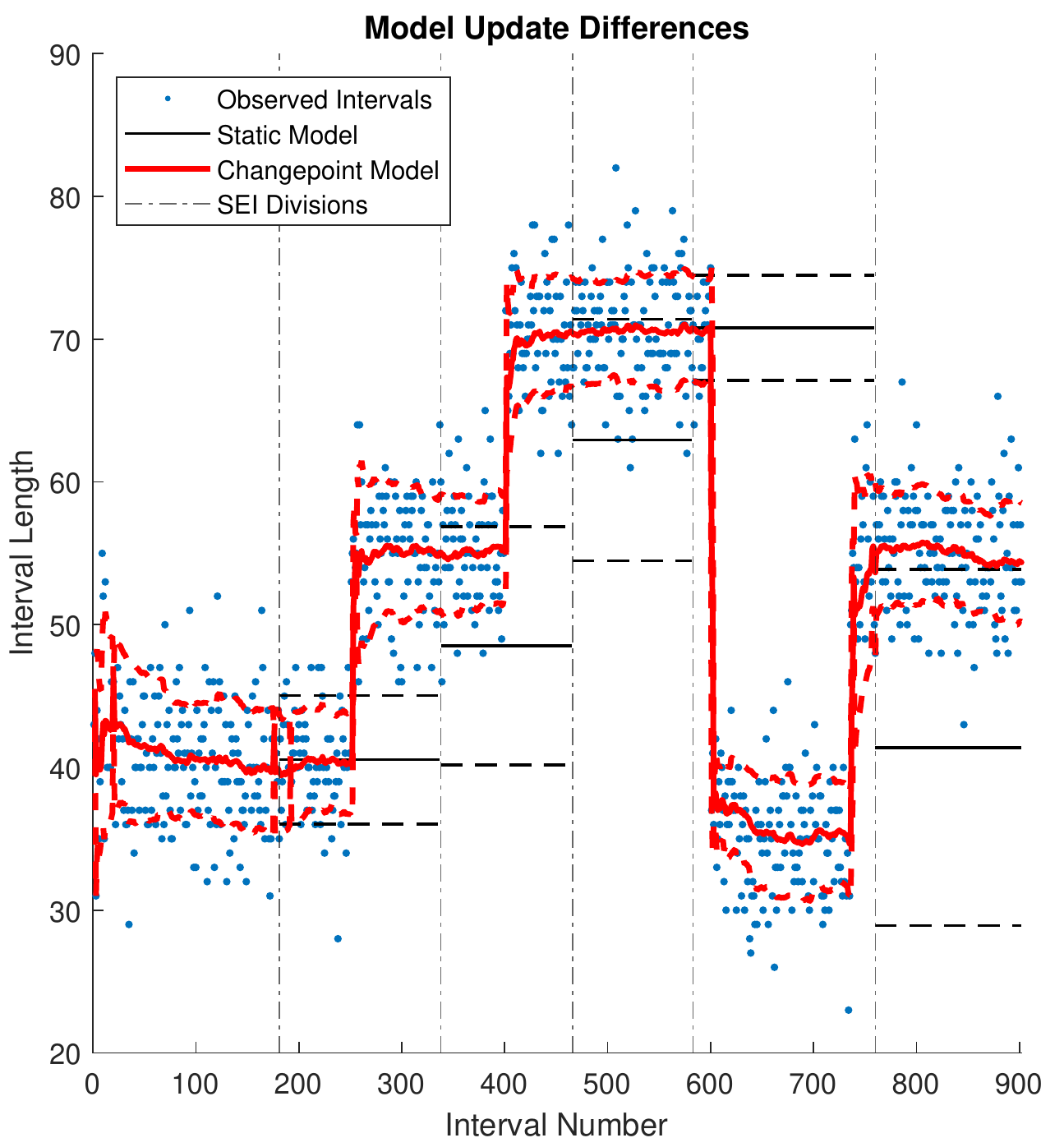}
    \caption{Example of changepoint modeling vs. periodic model updates at spectrum evaluation intervals (denoted by vertical lines). Idle intervals and models shown. Solid model lines denote estimated $\mu_{I_i}$ and dashed lines denote estimated $\mu_{I_i} \pm \sigma_{I_i}$. Note that spectrum evaluation intervals are evaluated by pulse number not interval number, so lines are not evenly dispersed in the interval domain.}
    \label{fig:my_label4}
\end{figure}

The sense and predict algorithm with changepoint detection performs significantly better than the original algorithm in the presence of changepoints. When each changepoint occurs, the modified algorithm is able to quickly adjust its models to accommodate the change in data, while the original is unable to do so until the next spectrum evaluation interval. Even though the static model periodically retrains its models, it often includes unrelated data from a previous partition. This leads to uncharacteristic sample estimates. The altered algorithm is also able to begin transmission on initialization much more quickly than the original, as it does not require a passive spectrum evaluation interval. An example of the differences in model update behavior is shown in Figure \ref{fig:my_label4}.

In the absence of changepoints, the altered algorithm is not able to perform as well as the original. This occurrence is primarily due to a lack of online threshold tuning.

Each sub-band in an environment is modeled and evaluated independently; as such, presented simulation results consider a single sub-band for the purposes of clarity without a loss of generality. The simulation environment used for testing draws busy and idle intervals from a probability distribution to present to the different versions of the sense and predict algorithm. At every time-step there is a constant probability $h$ that a changepoint will occur. If a changepoint does occur, the means of both busy and idle distributions will change. The magnitude of this change, $|\Delta|$, is drawn from its own probability distribution.

Specific algorithm performance is dependent on a great variety of parameters. Parameters for each sub-band include: parameters for ground truth busy and idle time duration probability distributions BD and ID, 
changepoint frequency $h$, and parameters for the changepoint magnitude distribution $|\Delta|$. Additional design parameters (where applicable) include collision error weighting $\alpha$, the length of the spectrum evaluation interval in pulses SEI, the length of the system's action-latency period in pulses $\Delta t$, the maximum run length parameter $L$, and the BOCD sensitivity parameter $\gamma$.

A table with sample simulation results is included in Table \ref{table_example}. The performance metric is the weighted error rate $\rho$, which is evaluated by a weighted combination of the missed opportunity rate $D$ and the collision rate $C$, using \eqref{rho}. The performance metrics of the original sense and predict algorithm are denoted by $D_O$, $C_O$, and $\rho_O$, while those of the altered algorithm employing the lognormal model are denoted by $D_L$, $C_L$, and $\rho_L$, and those of the altered algorithm employing the nonparametric model are denoted by $D_N$, $C_N$, and $\rho_N$.

\begin{table}[htbp]
\setlength{\tabcolsep}{4pt}
\renewcommand{\arraystretch}{1.3}
\caption{Sample simulation results comparing the performance of the unaltered sense and predict algorithm with the modified algorithm. }
\label{table_example}
\centering
\begin{tabular}
%
{|c|c|c|c|c|}
\hline
Test \# &$1$&$2$&$3$&$4$\\
\hline
BD& $\sim N(150,4)$& $\sim N(50,10)$& $\sim N(150,4)$& $\sim N(150,4)$\\
\hline
ID & $\sim N(150,4)$& $\sim N(50,10)$& $\sim N(150,4)$& $\sim N(150,4)$\\
\hline
$h$&$0.03$ &$0$&$0$&$0.03$\\
\hline
$|\Delta|$& $\sim N(40,10)$&-& - & $\sim N(40,10)$\\
\hline
SEI&$5000$&$5000$&$5000$&$5000$\\
\hline
$L$&$60$&$60$&$60$&$30$\\
\hline
$\gamma$&$60$&$60$&$60$&$60$\\
\hline
$\Delta t$&$5$&$5$&$5$&$5$\\
\hline
$\alpha$&$0.5$&$0.5$&$0.5$&$0.5$\\
\hline \hline
$C_O$&$0.2536$&$0.0984$&$0.0208$&$0.2293$\\
\hline
$D_O$&$0.2752$&$0.1001$&$0.0210$&$0.2895$\\
\hline
$\rho_O$&$0.2644$&$0.0993$&$0.0209$&$0.2594$\\
\hline
\hline
$C_L$&$0.0872$&$0.0990$&$0.0199$&$0.0847$\\
\hline
$D_L$&$0.1016$&$0.1006$&$0.0196$&$0.0980$\\
\hline
$\rho_L$&$0.0943$&$0.0998$&$0.0197$&$0.0913$\\
\hline
\hline
$C_N$&$0.1370$&$0.0964$&$0.0211$&$0.1187$\\
\hline
$D_N$&$0.1072$&$0.1055$&$0.0188$&$0.1119$\\
\hline
$\rho_N$&$0.1221$&$0.1010$&$0.0200$&$0.1153$\\
\hline

\end{tabular}
\end{table}

These results show that the introduction of changepoint detection provides a significant improvement to the algorithms performance when the underlying environmental model is allowed to change ($h=0.03$) while also maintaining the same level of performance under a consistent environmental model ($h=0$).

Test 1 shows algorithm performance in an environment with small variance in busy and idle distributions. Test 2 demonstrates the effect of large variations in busy and idle distributions even in the absence of changepoints. Test 3 demonstrates the affect of an absence of changepoints as compared to Test 1. Test 4 primarily demonstrates the small effect of the maximum run length parameter $L$ on the accuracy of the altered algorithms.

The unaltered algorithm is significantly affected both by the presence of changepoints and the variation of busy and idle distributions, while the altered versions of the algorithm are significantly affected by the variation of busy and idle distributions but affected much less significantly by the presence of changepoints. 

Further examination of parameter effects on system performance and thorough parameter tuning is not performed in these simulation results as system performance can be very strongly affected by the specific environment in which it operates. Parameter tuning is likely required when incorporating the altered sense and predict algorithm in a live spectrum sharing system.

\section{Conclusions}
In this work Bayesian online changepoint detection is applied to the sense and predict algorithm to increase its model accuracy in a changing environment. Changepoint detection is the process of identifying sudden changes in the distribution from which data is drawn. This detection allows an algorithm to include only relevant data when maintaining dynamic models of the environment, rather than including irrelevant data from a previous distribution.

As expected, the addition of changepoint detection to the sense and predict algorithm greatly increases its performance in the presence of changepoints. Informed knowledge of changepoint locations allows the system to more quickly and efficiently age out old and unrelated data. 
In the absence of changepoints, the altered versions of the algorithm maintain their performance, while the original algorithm shows significant performance degradation when changepoints are present. The performance of the altered algorithms is largely affected by the variance of busy and idle distributions.

This method of changepoint detection can likewise be implemented into other systems that maintain a model of the environment.

\end{document}